\documentclass[floatfix,aps,showpacs,twocolumn,superscriptaddress,
nofootinbib,preprintnumbers]
{revtex4}

\usepackage{graphicx}
\usepackage{dcolumn}

\newcommand{\ba}{\begin{eqnarray}}
\newcommand{\ea}{\end{eqnarray}}

\def\Frac#1#2{\frac{\displaystyle{#1}}{\displaystyle{#2}}}

\begin{document}

\preprint{FERMILAB-Pub-03/323-A} \preprint{LPT-ORSAY/04-13}

\title{Inflationary potentials yielding constant scalar perturbation
spectral indices}

\author{Alberto Vallinotto}
\affiliation{Physics Department, The University of Chicago, Chicago,
Illinois 60637-1433, USA, and \\
Dipartimento di Fisica, Politecnico di Torino, C.so Duca degli
Abruzzi 24, 10129 Torino, Italy. }

\author{Edmund J. Copeland}
\affiliation{Department of Physics and Astronomy, University of Sussex,
Brighton BN1 9QH, UK.}

\author{Edward W. Kolb}
\affiliation{Fermilab Astrophysics Center, Fermi National
Accelerator Laboratory, Batavia, Illinois, 60510-0500, USA, \\
and Department of Astronomy and Astrophysics, Enrico Fermi
Institute, \\ University of Chicago, Chicago, Illinois 60637-1433,
USA}

\author{Andrew R. Liddle}
\affiliation{Astronomy Centre, University of Sussex, Brighton BN1 9QH, UK.}

\author{Dani\`ele A. Steer}
\affiliation{Laboratoire de Physique Th\'eorique, B\^at. 210,
Universit\'e Paris XI, 91405 Orsay Cedex, France, and \\
F\'ed\'eration de Recherche APC, Universit\'e Paris VII, France.}

\date{\today}

\begin{abstract}
We explore the types of slow-roll inflationary potentials that result
in scalar perturbations with a {\em constant} spectral index, {\em
i.e.,} perturbations that may be described by a single power-law
spectrum over all observable scales.  We devote particular attention
to the type of potentials that result in the Harrison--Zel'dovich
spectrum.
\end{abstract}


\maketitle

\section{Introduction}
\label{sec:intro}

Inflation, a cornerstone of the modern framework for understanding
the  early universe \cite{guth81,lrreview}, predicts the initial
conditions for the formation of structure and the cosmic microwave
background (CMB) anisotropies.  During inflation, the primordial
scalar (density) and tensor (gravitational wave) perturbations
generated by quantum fluctuations are redshifted beyond the Hubble
radius, becoming frozen as perturbations in the background metric
\cite{muk81,hawking82,starobinsky82,guth82,bardeen83}.  However,
even when there is only one scalar field --- the {\it inflaton}
--- the number of inflation models proposed in the literature is
large \cite{lrreview}. Determination of the
properties of the scalar perturbations and tensor perturbations
from CMB and large-scale structure observations allows one to
constrain the space of possible inflation models
\cite{dodelson97,kinney98a,probes,wmapinf,barger,kkmr,liddleleach}.

It is often adequate to characterize inflationary perturbations in
terms of four quantities: the scalar and tensor power spectra,
$\mathcal{P}_{\mathcal{R}}$ and $\mathcal{P}_{g}$, and the scalar
and tensor spectral indices $n$ and $n_T$.  In this paper we focus
on the {\em scalar} spectral index which, unless explicitly
indicated otherwise, we refer to simply as the `spectral index'.
Successful inflation models predict $n$ close to $1$ (the
so-called Harrison--Zel'dovich spectrum), and $n$ typically has a
small scale dependence. The best data available to date, combining
the Wilkinson Microwave Anisotropy Probe \cite{wmap} and Sloan
Digital Sky Survey \cite{SDSS} data sets, indicate that the
evidence for anything other than a scale-invariant spectra is
marginal at best, with no evidence for significant running of the
scalar spectral index \cite{tegmarksloan}. Moreover, one of us has
recently argued that when information criteria are used to carry
out cosmological model selection based on the current data sets
available, then the best present description of cosmological data
uses a scale-invariant ($n=1$) spectrum \cite{liddleparameters}.
It therefore makes sense to be considering the inflationary
potentials associated with that spectrum.

It is known that inflaton potentials $V(\phi)=
\exp(-\alpha\phi)$ for constant $\alpha^2<2$ lead to perturbation
spectra that are exact power laws, {\it i.e}.~$n$ is a constant \cite{lucchin}.
However, there has not yet been a systematic analysis of the types
of inflaton potentials that yield constant $n$.
Here we take a first step in that direction, classifying those
potentials within the framework of the slow-roll approximation
\cite{SRref}.

In the next section the basic results employed to calculate the
properties of the perturbation spectrum using the slow-roll
parameterization of the inflaton potential are reviewed. In
Sec.~\ref{sec:method} two exact differential equations connecting the
potential and the field to the slow-roll parameters are derived
and the general method used to calculate all the relevant
cosmological quantities is outlined. In Sec.~\ref{sec:app} this method is
applied to the determination of the inflationary
potential yielding a $k$-independent density spectral index: both
the Harrison--Zel'dovich $(n=1)$ and the general $(n=1-2n_0^2)$
case are considered to lowest order and to next order in the
slow-roll parameter approximation. In Sec.~\ref{Sect:flow} the
flow of $\epsilon$ is examined to understand the number of
solutions that arise. The conclusions are contained in Sec.~\ref{sec:dis}.

\section{Review of Basic Concepts}
\label{sec:review}

\subsection{Inflationary Dynamics and Slow Roll Parameters}
\label{subsec:slow}

The dynamics of the standard Friedmann--Robertson--Walker (FRW)
universe driven by the potential energy of a single scalar field
-- the inflaton $\phi$ -- are usually expressed by the Friedmann
equation for flat spatial sections and by the energy conservation equation:
\ba
  H^2&=&\frac{8\pi}{3M_{{\rm p}}^2} \left[\frac{1}{2}\dot{\phi}^2+V(\phi)
  \right],
\label{Friedmann}
\\
 & & \ddot{\phi}+3H\dot{\phi}+V'(\phi)=0,
 \label{nrgconsfield}
 \ea
where $V(\phi)$ is the inflaton potential, $M_{{\rm p}}=G^{-1/2}$
the Planck mass and $H=\dot{a}/a$ the Hubble expansion parameter.
Once $V(\phi)$ is specified, the field dynamics are determined by
solving the coupled equations (\ref{Friedmann}) and
(\ref{nrgconsfield}). Often it is simplest to do this using the
Hamilton--Jacobi approach \cite{HJref} in which $H(\phi)$ is
considered the fundamental quantity to be specified.  Equations
(\ref{Friedmann}) and (\ref{nrgconsfield}) then become two
first-order equations:
\begin{eqnarray}
H'(\phi)^2 - 12 \pi H^2(\phi)/M_{{\rm p}}^2 & = & -32 \pi^2 V(\phi)/M_{{\rm
p}}^4 \,;\\
\dot{\phi} &=& -M_{{\rm p}}^2 H'(\phi)/4 \pi \,,
\end{eqnarray}
where $' \equiv d/d\phi$.
Whichever the method, once
the dynamics of the inflaton field is known, $a(t)$ is obtained by
integrating Eq.~(\ref{Friedmann}). Without any loss of generality
we assume that $\dot{\phi}>0$ during inflation. Here we use the
Hubble slow-roll parameters $\epsilon$, $\eta$ and $\xi^2$ as
defined in Ref.~\cite{LPB}
\begin{eqnarray}
\label{eps}
  \epsilon(\phi) &\equiv& \frac{3\dot{\phi}^2}{2}\left[ V(\phi)+
       \frac{\dot{\phi}^2}{2}
  \right]^{-1}=
  \frac{M_{{\rm p}}^2}{4\pi}{\left[\frac{H'(\phi)}{H(\phi)}\right]}^2, \\
\label{eta}
  \eta(\phi) &\equiv& -\frac{\ddot{\phi}}{H\dot{\phi}}
  =\frac{M_{{\rm p}}^2}{4\pi}\frac{H''(\phi)}{H(\phi)}, \\
\label{xi2}
  \xi^2(\phi) &\equiv& \frac{M_{{\rm p}}^4}{16\pi^2}
  \frac{H'(\phi)H'''(\phi)}{H^2(\phi)}.
\end{eqnarray}
The parameters $\eta$ and $\xi^2$ are the first terms in an
infinite hierarchy of slow-roll parameters, whose $l$-th member is
defined by
\begin{equation}\label{slrollparms}
  \lambda_H^l (\phi)\equiv \left(\frac{M_{{\rm p}}^2}{4\pi}\right)^l
  \frac{(H')^{l-1}}{H^l} \frac{d^{(l+1)}H(\phi)}{d\phi^{(l+1)}}.
\end{equation}
During slow-roll $\{\epsilon, \lambda_H^l  \} \ll 1$, and
inflation ends when $\epsilon = 1$. The potential and its
derivatives can be expressed as \textit{exact} functions of these
slow-roll parameters: up to second order in derivatives of $V$ one
has
\begin{eqnarray}
\label{V}
  V(\phi)&=&\frac{M_{{\rm p}}^2}{8\pi}H^2(3-\epsilon), \\
\label{dVdphi0}
  \frac{dV(\phi)}{d\phi}&=&-\frac{M_{{\rm
p}}}{2\sqrt{\pi}}H^2\sqrt{\epsilon}(3-\eta),\\
\label{d2Vdphi2}
  \frac{d^2V(\phi)}{d\phi^2}&=&H^2\left[ 3\epsilon+3\eta
  -(\eta^2+\xi^2)\right].
\end{eqnarray}

\subsection{A Hierarchy of Approximation Orders}
\label{subsec:hier}

As mentioned in the introduction, the observable quantities of
interest are the power spectrum $\mathcal{P}_{\mathcal{R}}$ of the
curvature perturbation $\mathcal{R}$ on comoving hypersurfaces and
the spectrum of gravity waves $\mathcal{P}_{g}$.  These define
$n(k)$ and $n_T(k)$ through
\begin{eqnarray}
\label{defnr}
  n(k)-1 & \equiv & \frac{d \ln \mathcal{P}_\mathcal{R}(k)}{d \ln k}, \\
\label{defnT}
  n_T(k)& \equiv & \frac{d \ln \mathcal{P}_{g}(k)}{d \ln k}.
\end{eqnarray}
As discussed in Ref.~\cite{Lidsey:1995np,Stewart:1993bc}, the
expressions for these quantities differ depending on the
approximation order assumed in the slow-roll expansion. The
approximation order is defined in general by considering how many
terms in a slow-roll parameter expansion of a generic expression
are retained, \textit{lowest-order} approximation corresponding to
retaining only the lowest-order term and \textit{next-order}
approximation corresponding to retaining terms up to the
next-to-lowest order term.

For the perturbation power spectra and spectral indices, the
lowest-order term is linear in the slow-roll parameters. To order
$l_0$, these expressions will contain the set of slow-roll
parameters $\{\epsilon, \lambda_H^l \}$ with $l=(1,2,..l_0)$ where
$\lambda_H^l$ is a term of order $l$. At \textit{next-order}
($l_0=2$), the expressions will contain the parameters
$\{\epsilon, \eta, \xi^2\equiv\lambda_H^2\}$ as well as all
second-order combinations thereof (namely $\epsilon^2, \eta^2$ and
$\eta \epsilon$). Hence, for \textit{order consistency}, whenever
an exact and an approximate expression are combined (as shall
often be the case below) the result is accurate only to the order
of the approximate expression, and the result must be expanded in
a power series of slow-roll parameters up to and including terms
of an overall degree consistent with the level of approximation
assumed.

Recalling Lidsey {\it et al.}~\cite{Lidsey:1995np}, it is then
possible to think of an infinite hierarchy of expressions for the
perturbation spectra and for the spectral indices. It is
unfortunate that, due to the complexity of the problem, only the
first two approximation orders are currently available in general: indeed, at
next-to-lowest order,
\begin{eqnarray}
\label{ASloword} \mathcal{P}_{\mathcal{R}}^{1/2}(k) & \simeq & 2
\left[1-\{(2C+1)\epsilon-C\eta\}\right]
   \left. \frac{H^2}{M_{{\rm p}}^2|H'|} \right|_{k=aH} \\
\label{ATloword}
 \mathcal{P}_{g}^{1/2}(k) & \simeq  & \frac{4}{\sqrt{\pi}}
    \left[1-\{(C+1)\epsilon\}\right] \left. \frac{H}{M_{{\rm p}}}
\right|_{k=aH},
   \\
    \label{nrloword}
  n(k)-1 & \simeq & -4\epsilon + 2\eta-\{ 8(C+1)\epsilon^2 \nonumber\\
  & & -(6+10C)\epsilon\eta+2C\xi^2\}, \\
\label{ngloword}
  n_T(k) & \simeq &
  -2\epsilon-\{2\epsilon^2(3+2C)-4(1+C)\epsilon\eta\},
\end{eqnarray}
where $C\simeq-0.73$ \cite{Lidsey:1995np,Stewart:1993bc}. As in
Ref.~\cite{Lidsey:1995np}, the symbol ``$\simeq$'' is used to indicate
that the results are accurate up \textit{to the order of
approximation assumed}. The lowest-order results are obtained by
setting all the terms in curly brackets to zero.

\section{The Parametrization Method}
\label{sec:method}

We now focus on the case of constant $n(k)$.  To any order $l_0$ in the
slow-roll
approximation, imposing $k$-independence of $n(k)$ endows the
problem with the additional set of $(l_0-1)$ relations
\begin{equation}\label{nn=0}
  \frac{d^i n(k)}{d (\ln k)^i}=0, \qquad  i=1,\ldots,(l_0-1).
\end{equation}
Therefore, since there are $l_0+1$ slow-roll parameters at this
order, the conditions (\ref{nn=0}) together with the constancy of
$n(k)$ mean that only one of those is independent: throughout the
rest of this paper we take it to be $\epsilon$.  As we show in
this section, it is then possible to determine $\phi(\epsilon)$
and $V(\phi)$ to this order.

The method is the following.  First we derive two exact
differential equations for $\phi$ and $V$ which, as we shall see
below, only contain the slow-roll parameters $\eta$ and
$\epsilon$.  Then, at a given order $l_0$, we impose the
conditions given in Eq.~(\ref{nn=0}) which yield $\eta(\epsilon)$.
As a result the two differential equations can be integrated to
obtain $V(\epsilon)$ and $\phi(\epsilon)$ correct to order $l_0$.
Finally, provided $\phi(\epsilon)$ can be inverted, we can obtain
$V(\phi)$. This will be done in the next section where we also solve for all
the dynamics of the problem, namely $H(\phi)$, $a(t)$ and
$\phi(t)$.

{}From Eq.(\ref{eps}) it is straightforward to obtain
\begin{equation}
\label{depsdphi} \frac{d\epsilon}{d\phi}=
  \frac{2 M_{{\rm p}}^2}{4\pi} \left[
  \frac{H'H''}{H^2}-{\left(\frac{H'}{H}\right)}^3\right],
\end{equation}
which, together with the definitions of $\epsilon$ and $\eta$,
yields the \textit{exact} differential equation
\begin{equation}\label{depsdphi2}
 \frac{d\epsilon}{d\phi}= \frac{4 \sqrt{\pi}}{M_{{\rm p}}}
 \sqrt{\epsilon}(\epsilon-\eta) \,.
\end{equation}
Once $\eta(\epsilon)$ is specified, integration of this
equation yields $\phi(\epsilon)$.

Also, Eqs.~(\ref{dVdphi0}) and (\ref{depsdphi2}) give
\begin{equation}\label{dVdeps0}
    \frac{dV}{d\epsilon}=\frac{dV}{d\phi}\frac{d\phi}{d\epsilon}=-\frac{M_{{\rm
p}}^2
    H^2}{8\pi}\left[ \frac{3-\eta}{\epsilon-\eta} \right],
\end{equation}
which, divided by Eq.~(\ref{V}), produces the following
\textit{exact} differential equation, useful because it is independent
of the Hubble parameter:
\begin{equation}\label{dlnVdeps}
\frac{1}{V}\frac{dV}{d\epsilon}=\frac{3-\eta}{(\eta-\epsilon)(3-\epsilon)}=\frac
{1}{\epsilon-3}+\frac{1}{\eta-\epsilon}.
\end{equation}
Given $\eta(\epsilon)$, Eq.~(\ref{dlnVdeps}) can be
integrated to give
\begin{equation}\label{Veps}
    V(\epsilon)=V_0 |3-\epsilon| \exp \left[ \int
    \frac{d\epsilon}{\eta(\epsilon)-\epsilon}\right],
\end{equation}
where $V_0$ is the integration constant which can be obtained from the observed
perturbation amplitude. Finally from Eq.~(\ref{V}) the following expression
for $H$ can be obtained
\begin{equation}\label{Heps}
    H^2(\epsilon)=\frac{8\pi V_0}{M_{{\rm p}}^2}\exp \left[ \int
    \frac{d\epsilon}{\eta(\epsilon)-\epsilon}\right].
\end{equation}
As noted in the previous section, once the integrations in
Eqs.~(\ref{Veps}) and (\ref{Heps}) have been carried out, order
consistency requires that the resulting expressions are expanded
in powers of $\epsilon$ and only terms up to and including order
$l_0$ are kept.

Once the expressions for $V(\epsilon)$ and $\phi(\epsilon)$ have
been computed, it is then possible to determine all the other
relevant cosmological quantities. Eq.~(\ref{Heps}) together with
the expression for $\epsilon(\phi)$ gives $H(\phi)$ to the given
order $l_0$. This, together with the equation obtained for
$V(\phi)$ then enables $\phi(t)$ to be calculated using
Eq.~(\ref{nrgconsfield}).\footnote{Once again, note that the
conservation equation must be truncated to the correct order $l_0$
in the approximation scheme.} Once this step is carried out, the
time evolution of the Hubble parameter can be derived -- either
using Eq.~(\ref{nrgconsfield}) or the solution of Eq.~(\ref{Heps})
-- and its integration then yields the dynamics of the scale
factor $a(t)$.

Before turning to the specific cases of constant spectral index, it is worth
commenting on the apparently singular case of $\eta=\epsilon$. This is
nothing other than the usual exact power-law inflation model and is perfectly
regular. From  Eq.~(\ref{depsdphi2}), we see that in this case the solution is
$\epsilon=\epsilon_0$, a constant independent of $\phi$. Substituting this value
into Eqs.~(\ref{eps}) and (\ref{V}) we obtain
\begin{eqnarray}\label{Hpowerlaw}
H &=& {\sqrt{8\pi V_0} \over M_{{\rm p}}} \exp \left[-{2\sqrt{\pi \epsilon_0}
\phi \over
M_{{\rm p}}} \right] \\
V &=& V_0 (3-\epsilon_0) \exp \left[-{4\sqrt{\pi \epsilon_0} \phi \over M_{{\rm
p}}}
\right].
\end{eqnarray}
Substituting this into the Friedmann equation,  Eq.~(\ref{Friedmann}), we
obtain $\phi(t)$ through
\begin{equation}
\label{Phipowerlaw}
\sqrt{8\pi V_0} {\epsilon_0 t \over M_{{\rm p}}} = \exp \left[{2\sqrt{\pi
\epsilon_0}
\phi \over M_{{\rm p}}} \right] \,.
\end{equation}
Hence in Eq.~(\ref{Hpowerlaw}) we find $a(t) \sim t^p$ where
$p=1/\epsilon_0$, the usual power-law inflation result.

Finally, we note that it is also possible to address the present
problem using the definitions of the slow-roll parameters in the
expression for the spectral index to obtain a differential
equation for $H(\phi)$ \cite{beato}. While at lowest-order this
approach yields results which are equivalent to the ones derived
in the next section,\footnote{It is straightforward to show that
the condition $\eta=W\epsilon$ for $W \neq 1$ is solved by
$H(\phi)=A+B\phi^{1/(1-W)}$.} the differential equation arising at
next-order does not seem to allow an analytical solution and in
that case the parametrization method outlined above proves to be
preferable.

\section{Applications}
\label{sec:app}

In this section the method outlined above is applied to the
determination of the inflationary potentials which yield a
$k$-independent spectral index. Two cases will be considered: the
Harrison--Zel'dovich power spectrum, and the case of a
$k$-independent spectral index not equal to unity. For each case,
both lowest-order and next-order approximation results will be
derived.

\subsection{The Harrison--Zel'dovich Case}
\label{subsec:n1}

\subsubsection{Lowest-order approximation}
\label{Sect:Hzlo}

Imposing $n(k)=1$ in the lowest-order expression for the spectral
index, Eq.~(\ref{nrloword}), yields
\begin{eqnarray}\label{HZloconstraint}
  \eta(\epsilon) \simeq 2\epsilon.
\end{eqnarray}
Thus Eqs.~(\ref{depsdphi2}) and (\ref{Veps}) become
\ba
\label{Hzlo:deps}
  \frac{d\epsilon}{d\phi} &\simeq& -\frac{4 \sqrt{\pi}}{M_{{\rm p}}}
 \epsilon^{3/2},
\\
\label{Hzlo:dlnV}
  \frac{d \ln V}{d\epsilon} &\simeq &
  \frac{1}{\epsilon-3}+\frac{1}{\epsilon},
\ea
which can be integrated immediately, giving
\ba
\label{Hzlo:phieps}
  \phi(\epsilon) &\simeq & \frac{M_{{\rm p}}}{2\sqrt{\pi\epsilon}},
\\
\label{Hzlo:Veps}
  V(\epsilon) & \simeq & V_0 (3-\epsilon) \epsilon \simeq V_0
  3\epsilon ,
\ea
and hence
\begin{equation}\label{V3}
  V(\phi)\simeq V_0 \frac{3 M_{{\rm p}}^2}{4 \pi \phi^2}.
\end{equation}
Eq.~(\ref{Heps}) then yields
\begin{equation}\label{Hzlo:Heps}
    H^2(\phi)\simeq\frac{8\pi V_0}{M_{{\rm p}}^2} \epsilon \simeq \frac{2
V_0}{\phi^2}\,,
\end{equation}
and the constant $V_0$ can be read off from the lowest-order version of
Eq.~(\ref{ASloword}) as
\begin{equation}
V_0 \simeq \frac{M_{{\rm p}}^4}{8} \, {\cal P}_{\cal R} \,.
\end{equation}

This, together with the expression for $V'(\phi)$, can then be
used in the Friedmann equation which becomes
\begin{equation}\label{Hzlo:Friedmann}
    \phi^2 \dot{\phi}\simeq\frac{\sqrt{2 V_0}M_{{\rm p}}^2}{4 \pi},
\end{equation}
so that
\begin{equation}\label{phioft2}
  \phi(t) \simeq \phi_0 \ (t/t_0)^{1/3},
\end{equation}
where $\phi_0^3t_0^{-1}=3\sqrt{2V_0}M_{{\rm p}}^2/4\pi$.
Eq.~(\ref{Hzlo:phieps}) can then be used to compute the dynamics
of the slow-roll parameter
\begin{equation}\label{epsfctt}
  \epsilon(t) \simeq \frac{M_{{\rm p}}^2}{4\pi\phi_0^2}\left(t/t_0\right)^{-2/3}.
\end{equation}
Finally, the time evolution of the Hubble parameter and of the
scale factor are given by:
\begin{eqnarray}\label{Hfctt}
  H(t) & \simeq & H(t_0)\left(t/t_0\right)^{-1/3} ,  \nonumber\\
  \frac{a(t)}{a(t_0)} & \simeq & \exp  \left\{
   \frac{\sqrt{8 V_0}}{3\phi_0 t_0^{-1}}
   \left[ \left(\frac{t}{t_0}\right)^{2/3}-1 \right] \right\} .
\end{eqnarray}

Let us now recall the work of Barrow and Liddle on
\textit{intermediate inflation} \cite{Barrow:1993zq}.
Though the present work differs in spirit from
that paper (which starts by postulating a specific
dynamics and then goes on
to derive the corresponding potential), the two approaches share a
common point, as we now outline. In Ref.~\cite{Barrow:1993zq} the scale
factor is assumed to take the form
\begin{equation}
\label{Barrow1}
  a(t)=\exp \left( At^f \right),
\end{equation}
with $0<f<1$, $A>0=$ constants. The authors then prove that this
is an exact solution of the `intermediate' inflation potential
\begin{equation}\label{Vbarrow}
  V(\phi)=\frac{8A^2}{(\beta+4)^2}\left[ \frac{(2A\beta)^{1/2}}{\phi}
  \right]^{\beta} \left[ 6- \frac{\beta^2}{\phi^2} \right],
\end{equation}
where $\beta=4(f^{-1} -1)$, and that it is also a solution
\textit{in the slow-roll approximation} for the potential
\begin{equation}\label{Vbarrow2}
  V(\phi)=\frac{48A^2}{(\beta+4)^2}\left[
  \frac{(2A\beta)^{1/2}}{\phi}
  \right]^{\beta}.
\end{equation}

To see how the present results relate to the ones reported in
Ref.~\cite{Barrow:1993zq}, we first quote the expressions for the
slow-roll parameters obtained in the intermediate inflation case:
\begin{equation}
\label{epsbarrow} \epsilon = \frac{\beta^2}{2\phi^2} \,; \quad
\eta = \left(1+\frac{\beta}{2}\right)\frac{\beta}{\phi^2} \,.
\end{equation}
Exploiting Eq.~(\ref{epsbarrow}), the equation
for the exact intermediate inflation potential can be recast in
the form
\begin{equation}\label{Vbarrow3}
  V(\phi) =\frac{16A^2}{(\beta+4)^2}\left[
  \frac{(2A\beta)^{1/2}}{\phi}
  \right]^{\beta} [3- \epsilon(\phi)].
\end{equation}
Now, we can think of this expression
as a function of the slow-roll parameter $\epsilon$ instead of the
field $\phi$. In this perspective, neglecting the $\epsilon$ in
the $(3-\epsilon)$ factor is the same as saying that
\textit{lowest-order} slow-roll approximation is assumed and that
by order consistency one should retain only the lowest-order term
arising from $\phi^{-\beta}(\epsilon)$. In other words, the
$\epsilon$ appearing in the $(3-\epsilon)$ factor will generate
terms of higher order, all of which can be consistently neglected
in a lowest-order calculation.

Note furthermore that imposing the $n(k)=1$ condition in the form
consistent with the lowest-order approximation (that is, $\eta=2
\epsilon$) and using Eq.~(\ref{epsbarrow})
yields $\beta=2$ and $f=2/3$. This is consistent with the previous
calculation,  since inserting this value of $\beta$ into Eq.~(\ref{Vbarrow2})
produces an expression for the inflaton potential
analogous to Eq.~(\ref{V3})
\begin{equation}\label{V3'}
  V(\phi) \sim \frac{3}{\phi^2},
\end{equation}
thus showing that the present analysis and the one carried out by
Barrow and Liddle in Ref.~\cite{Barrow:1993zq} agree on the lowest-order
potential able to produce a Harrison--Zel'dovich density
power spectrum.

\subsubsection{Next-order approximation}
\label{Sect:Hzno}

As discussed at the beginning of Sec.~\ref{sec:method}, the two
conditions given in Eq.~(\ref{nn=0}) must now be imposed in order
to determine $\eta(\epsilon)$.  The first condition is simply
obtained from Eq.~(\ref{nrloword}): imposing $n(k)=1$ at
next-order gives
\begin{equation}\label{neq1}
  4\epsilon - 2\eta +8
  (C+1)\epsilon^2-(6+10C)\epsilon \eta +2C\xi^2 \simeq 0.
\end{equation}
The second condition, $dn /d\ln k = 0$, yields
\cite{Lidsey:1995np}
\begin{equation}\label{nprimeq0}
-2\xi^2-8\epsilon^2+10\epsilon\eta \simeq 0.
\end{equation}
These expressions then allow us to solve for $\xi^2$ and $\eta$ as
functions of $\epsilon$, giving
\begin{eqnarray}
  \eta(\epsilon)& \simeq &\frac{2\epsilon+4\epsilon^2}{3\epsilon+1}
  \simeq 2\epsilon-2\epsilon^2, \nonumber \\
  \xi^2(\epsilon)&\simeq &\frac{6\epsilon^2+8\epsilon^3}{3\epsilon+1}
  \simeq 6\epsilon^2.
  \label{etaeps}
\end{eqnarray}
Eqs.~(\ref{depsdphi2}) and (\ref{Veps}) become
\begin{eqnarray}
  \frac{d\epsilon}{d\phi} &\simeq & -\frac{4 \sqrt{\pi}}{M_{{\rm p}}}
 \sqrt{\epsilon}\;\frac{\epsilon(\epsilon+1)}{3\epsilon+1},
\qquad\label{Hzno:deps}
\\
\label{Hzno:dlnV}
  \frac{d \ln V}{d\epsilon} &\simeq &
\frac{1}{\epsilon-3}+\frac{3\epsilon+1}{\epsilon(\epsilon+1)}.
\end{eqnarray}
These can be integrated exactly to yield
\begin{eqnarray}
  \phi(\epsilon) &\simeq  & \frac{M_{{\rm p}}}{2\sqrt{\pi}} \left[
\frac{1}{\sqrt{\epsilon}}
     - 2 \tan^{-1}(\sqrt{\epsilon}) \right] \nonumber \\
     && \simeq \frac{M_{{\rm p}}}{2\sqrt{\pi}}\left(
\frac{1}{\sqrt{\epsilon}}-2\sqrt{\epsilon}\right),\label{Hzno:phieps}
\\
\label{Hzno:Veps}
  V(\epsilon) &\simeq & V_0 \epsilon
  (3-\epsilon)(1+\epsilon)^2 \simeq V_0 (3\epsilon+5\epsilon^2).
\end{eqnarray}

\begin{figure}
\includegraphics[width=0.45\textwidth]{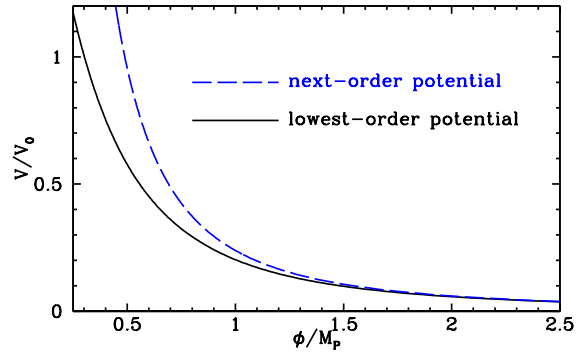}
\caption{\label{fig:2}Potentials giving the Harrison--Zel'dovich
density spectral index, computed to lowest-order approximation and
to next-order approximation.}
\end{figure}

In this case it is neither straightforward nor very enlightening
to obtain an explicit expression for the potential as a function
of the field. Numerically, however, we can determine $V(\phi)$
from Eqs.~(\ref{Hzno:phieps}) and (\ref{Hzno:Veps}). The result
is plotted in Fig.~\ref{fig:2} together with the lowest-order
result.

\subsection{General power-laws}
\label{Sect:n0}

Having determined the inflationary potential generating a
Harrison--Zel'dovich
spectrum, in this Section we consider the more general case for
which
\begin{equation}\label{defn0}
  n(k)=1-2n_0^2 \qquad \forall k.
\end{equation}
We focus primarily on the $n_0^2>0$ case: the results for
$n_0^2<0$ are obtained by analytic continuation, with some care
being taken over the number of solutions available in that case.

\subsubsection{Lowest-order approximation}
\label{subsec:LO}

\begin{figure}
\includegraphics[width=0.45\textwidth]{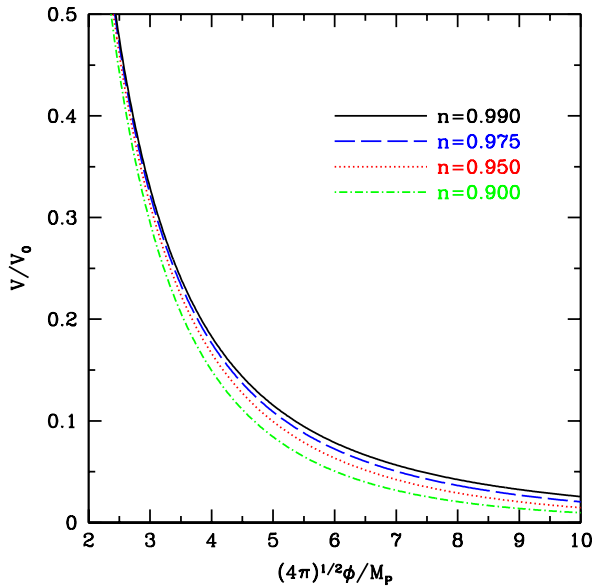}
\caption{\label{fig:3}Four potentials computed to lowest order,
yielding density perturbation spectral indices of 0.9, 0.95,
0.975, 0.99.}
\end{figure}

Inserting the lowest-order expression for $n(k)$, Eq.~(\ref{nrloword}), into
Eq.~(\ref{defn0}), gives
\begin{equation}\label{n0lo:eta}
  \eta(\epsilon) \simeq 2\epsilon-n_0^2,
\end{equation}
so that Eqs.~(\ref{depsdphi2}) and (\ref{Veps}) become
\begin{eqnarray}\label{n0lo:depsdphi}
    \frac{d\epsilon}{d\phi}& \simeq &\frac{4\sqrt{\pi}}{M_{{\rm p}}}\sqrt{\epsilon}
    (n_0^2-\epsilon),\\
  \frac{d \ln V}{d\epsilon} &\simeq & \frac{1}{\epsilon-3}+\frac{1}{\epsilon-n_0^2}.
\label{dlnVn0lo}
\end{eqnarray}
Let's first consider the $n_0^2>0$ case. Depending on whether
$\epsilon>n_0^2$ or $\epsilon<n_0^2$, integration of Eq.~(\ref{n0lo:depsdphi})
above yields
\begin{equation}
\phi(\epsilon) \simeq \frac{M_{{\rm p}}}{2 n_0 \sqrt{\pi}}  \left\{
 \begin{array}{cc}
\coth^{-1} \left(
  \sqrt{\epsilon/{n_0^2}}\right)  &  \qquad (\epsilon>n_0^2) \\
 \tanh^{-1} \left(
  \sqrt{\epsilon/{n_0^2}}\right)&  \qquad (\epsilon < n_0^2)\\
\end{array}
\right.
 \label{phiepsn0}
\end{equation}
Similarly, integration of Eq.~(\ref{dlnVn0lo}) gives
\begin{equation}
  V(\epsilon)\simeq  V_0 (3-\epsilon)|\epsilon-n_0^2|
   \simeq \pm V_0 [\epsilon(3+n_0^2)-3n_0^2],
\end{equation}
where upper (lower) sign refers to the $\epsilon>n_0^2$
$(\epsilon<n_0^2)$ case. Combining these results produces
\begin{equation}
V(\phi)  \simeq V_0 n_0^2\left\{
 \begin{array}{cc}
 -3 + (3+n_0^2)\coth^2\left( \Frac{2n_0\sqrt{\pi}}{M_{{\rm p}}} \phi
  \right) &  (\epsilon>n_0^2) \\
 3 -  (3+n_0^2)\tanh^2\left( \Frac{2n_0\sqrt{\pi}}{M_{{\rm p}}} \phi
  \right)  &  (\epsilon < n_0^2)\\
\end{array}
\right.
\end{equation}
Examples of such potentials for $\epsilon>n_0^2$ are illustrated
in Fig.\ \ref{fig:3}.

When $n_0^2<0$, the corresponding lowest-order results for $V(\epsilon)$ and
$\phi(\epsilon)$ are given by
\begin{equation}
  V(\epsilon) \simeq V_0 \left[\epsilon(3+n_0^2)-3 n_0^2\right],
  \label{eq:Ve2}
\end{equation}
and
\begin{equation}\label{phiepsn0lo}
    \phi(\epsilon) \simeq \frac{M_{{\rm p}}}{2 \sqrt{\pi|n_0^2|}} \tan^{-1} \left(
  \sqrt{\frac{\epsilon}{|n_0^2|}}\right).
\end{equation}
Inverting Eq.~(\ref{phiepsn0lo}) we obtain
\begin{equation}
  V(\phi) \simeq V_0 |n_0^2| \left[(3+n_0^2) \tan^2\left( \frac{2 \phi \sqrt{\pi
    |n_0^2|}}{M_{{\rm p}}}\right)- 3 \right],
\end{equation}
where now only one solution exists because $\epsilon-n_0^2>0$.

At this point it seems rather puzzling that there are two
different solutions for the potential when $n_0^2>0$, and only one
when $n_0^2<0$. In Sec.~\ref{Sect:flow} it will be shown that the
reason for this is related to the behavior that Eq.~(\ref{n0lo:depsdphi})
exhibits as a function of the initial value
of the slow-roll parameter, $\epsilon_0$.

\subsubsection{Next-order approximation}
\label{subsec:NOapp}

First it is necessary to express the slow-roll parameters $\eta$
and $\xi^2$ as functions of $\epsilon$ and $n_0^2$. At next-order
the condition (\ref{defn0}) gives
\begin{equation}\label{Bneq1}
  4\epsilon - 2\eta +8
  (C+1)\epsilon^2-(6+10C)\epsilon \eta +2C\xi^2 \simeq 2n_0^2.
\end{equation}
On imposing the condition $dn(k)/ d\ln k=0$ we find
\begin{eqnarray}
  \eta(\epsilon) & \simeq & \frac{2\epsilon+4\epsilon^2-n_0^2}{3\epsilon+1}
\\
 & & \simeq  -n_0^2+(2+3n_0^2)\epsilon-(2+9n_0^2)\epsilon^2,
   \nonumber \\
  \xi^2(\epsilon) & \simeq & \frac{6\epsilon^2+8\epsilon^3-5n_0^2\epsilon}
    {3\epsilon+1} \\
   &  & \simeq  -5n_0^2\epsilon+(6+15n_0^2)\epsilon^2,
\nonumber
\end{eqnarray}
so that Eqs.~(\ref{depsdphi2}) and (\ref{Veps}) in this case take
the form
\ba \label{Bdepsdphi4a}
 \frac{d\phi}{d\epsilon} & \simeq & -\frac{M_{{\rm p}}}{4 \sqrt{\pi}}
\frac{1}{ \sqrt{\epsilon}}\frac{3\epsilon+1}{\epsilon^2 + \epsilon-n_0^2},
\\
\label{dlnVdepsn0no}
    \frac{d\ln
V}{d\epsilon}&\simeq &\frac{1}{\epsilon-3}+\frac{3\epsilon+1}{\epsilon^2+\epsilon-n_0^
2}.
\ea
To solve these equations, let $a$ and $b$ be the two roots of
$\epsilon^2 + \epsilon-n_0^2 = 0 $ so that
\begin{equation}\label{n0no:abdelta}
    2a=-1-\delta \; , 2b=-1+\delta \; \; \; {\rm with} \; \; \;
\delta =\sqrt{1+4n_0^2}.
\end{equation}
Furthermore we assume  $0< n_0^2 \ll 1 $, so that $a \simeq
-(1+n_0^2)<0$ and $b \simeq n_0^2 >0$. Using
\begin{equation}\label{n0no:decomposition}
    \frac{3\epsilon+1}{\epsilon^2+\epsilon-n_0^2} =
    \frac{p_+}{\epsilon -a} +\frac{p_-}{\epsilon-b} \; \; \; {\rm
    with} \; \; \; p_{\pm} = \frac{(3 \pm \delta^{-1} )}{2}
\end{equation}
one can integrate Eq.~(\ref{Bdepsdphi4a}) to find, in the cases
$\epsilon> b \simeq n_0^2$ and $\epsilon<b \simeq n_0^2$
respectively,
\begin{equation}\label{n0no:phieps}
\phi(\epsilon) \simeq \frac{M_{{\rm p}}}{2\sqrt{\pi}}  \left\{
 \begin{array}{c}
-\Frac{p_+}{\sqrt{|a|}} \tan^{-1} \sqrt{\Frac{\epsilon}{|a|}}
+ \Frac{p_-}{\sqrt{b}} \coth^{-1}  \sqrt{\Frac{\epsilon}{b}}  \\
-\Frac{p_+}{\sqrt{|a|}} \tan^{-1} \sqrt{\Frac{\epsilon}{|a|}}
+ \Frac{p_-}{\sqrt{b}} \tanh^{-1}  \sqrt{\Frac{\epsilon}{b}} \\
\end{array}
\right.
\end{equation}
Finally, integration of Eq.~(\ref{dlnVdepsn0no}) yields
\begin{equation}\label{n0no:Veps}
    V(\epsilon) \simeq V_0 (3-\epsilon)
    \left|\epsilon-a\right|^{p_+}
    \left|\epsilon-b\right|^{p_-}
\end{equation}
As in Sec.\ \ref{Sect:Hzno} the potential and the field have been
successfully parametrized with respect to $\epsilon$: they can be
inverted numerically to find $V(\phi)$.

\section{The flow of $\epsilon$}\label{Sect:flow}

As was pointed out in Sec.~\ref{Sect:n0}, it is interesting that
more than one solution arises in the general power-law case. To further
explore the reason for this, it is necessary to consider again the
evolution of $\epsilon(\phi)$ given by Eq.~(\ref{n0lo:depsdphi}),
keeping in mind that without loss of generality $\dot{\phi}>0$ is
assumed.

\subsection{The $n_0^2>0$ case}

\begin{figure}
\includegraphics[width=0.45\textwidth]{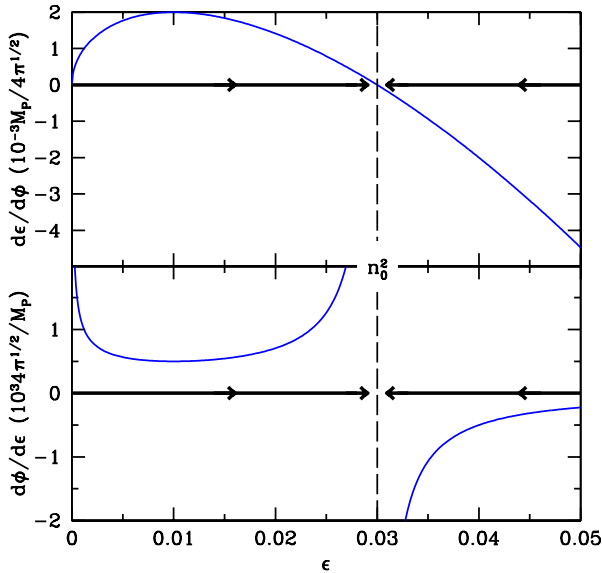}
\caption{\label{fig:4} The values of $d\epsilon/d\phi$ and
$d\phi/d\epsilon$ for an assumed value of $n_0^2=0.03$. Notice
that the sign of the derivatives implies that for $\epsilon
\rightarrow n_0^2$ the value of the field tends toward infinity.}
\end{figure}

From Fig.\ \ref{fig:4}, which shows $d\epsilon/d\phi$ as function
of $\epsilon$, it is possible to note that $d\epsilon/d\phi$ is
positive for $\epsilon<n_0^2$ and is negative for
$\epsilon>n_0^2$. One can see that if $\epsilon_0$, the initial
value of $\epsilon$, is smaller than $n_0^2$, then the slow-roll
parameter $\epsilon$ will increase toward $n_0^2$, while if the
initial value $\epsilon_0$ is greater than $n_0^2$, then
$\epsilon$ will decrease toward $n_0^2$. In the $n_0^2>0$ case,
then, independent of its initial value $\epsilon_0$, $\epsilon$
will tend toward the point $\epsilon=n_0^2$.

We have already seen that if $\epsilon = \eta$, then $\epsilon$ is a constant
given by $\epsilon_0=n_0^2$,
and that this fixed point corresponds to power-law
inflation generating a $k$-independent density spectral index
given by $n(k)=1-2n_0^2$. This result also allows one to reconcile
the apparent contradictory requirements for the generation of a
Harrison--Zel'dovich power spectrum stemming from the lowest-order
slow-roll approximation condition, $\eta=2\epsilon$, and by
power-law inflation definition $\epsilon=\eta=\xi=\cdots =n_0^2$.
One can see once again that a Harrison--Zel'dovich power spectrum
can be generated by power-law inflation in the limit $n_0^2
\rightarrow 0$ (i.e. $p \rightarrow \infty$), which corresponds to
pure de Sitter expansion \cite{Lidsey:1995np}.

Turning our attention to the case $\epsilon_0 \neq n_0^2$, it is
easier to consider the derivative of $\phi$ with respect to
$\epsilon$,
\begin{equation}\label{dphideps}
  \frac{d\phi}{d\epsilon}
  \simeq \frac{M_{{\rm p}}}{4\sqrt{\pi}}\frac{1}{\sqrt{\epsilon}(n_0^2-\epsilon)} ,
\end{equation}
which is also shown in Fig.\ \ref{fig:4}.
The interesting feature here is that the point
$\epsilon=n_0^2$ represents an asymptote of $d\phi/d\epsilon$:
integrating it on either side with $\epsilon \rightarrow n_0^2$
yields a logarithmically-diverging field. This necessarily implies
that the value of the field, parametrized by $\epsilon$, will tend
to infinity while $\epsilon$ tends toward $n_0^2$. Remembering
that Eq.~(\ref{n0lo:depsdphi}) is integrated to yield
$\phi(\epsilon)$, it is then possible to note that the three
distinct regions $\epsilon<n_0^2$, $\epsilon=n_0^2$ and
$\epsilon>n_0^2$ will give rise to three different dynamical
behaviors for $\phi$, which, once inserted in the expression for
$V(\epsilon)$, are able to produce the same density perturbation
spectral index. The apparent puzzle that arose at the end of Sec.\
\ref{subsec:LO} has therefore been solved: there are in fact two
potentials, and both their domains are $\phi \in [0,\infty[$. It
is now possible to understand that each one of them -- together
with power law inflation -- is able to generate the desired power
spectrum, depending on the initial condition chosen for the
slow-roll parameter.

\subsection{The $n_0^2\leq0$ case}

The cases $n_0^2=0$ and $n_0^2<0$ are similar. From Eq.~(\ref{n0lo:depsdphi}) we
see that,
independent of $\epsilon_0$, the value of $\epsilon$ will tend
toward zero as inflation proceeds. In the $n_0^2<0$ case the
solution derived Sec.\ \ref{Sect:n0} is the only one available,
while in the special case $n_0^2=0$ (Harrison--Zel'dovich) it is
possible to claim that two different inflationary potentials will
be able to generate such a power spectrum: the flat one giving
rise to the classical de Sitter expansion, and the one derived in
Sec.\ \ref{Sect:Hzlo}, whose first term is proportional to
$\phi^{-2}$.

\section{Discussion}
\label{sec:dis}

The analysis that has been carried out shows that inflaton
potentials yielding the Harrison--Zel'dovich flat spectrum can be
determined to \textit{lowest-order} and \textit{next-order}
approximation in the slow-roll parameters. Similarly, potentials
producing a $k$-independent spectral index slightly different from
unity have been derived to \textit{lowest-order} and to
\textit{next-order}.

It is also possible to speculate that the same procedure can be
carried out to any order of expansion in the slow-roll parameters.
This is because the implications of the spectral index
$k$-independence are not as trivial as they may seem at first
glance. Notice in fact that every time a higher approximation
order is assumed, new slow-roll parameters will appear in the
expression for the spectral index: going from lowest-order to
next-order, for example, $\xi^2$ was introduced. This is hardly surprising,
though, because these new
parameters just correspond to higher derivatives of $V(\phi)$ or
$H(\phi)$ (whatever is the degree of freedom chosen to express the
slow-roll parameters) and a higher order treatment necessarily
needs to take into account more derivative terms of the
potential. However, the requirement of the spectral index to be
$k$-independent implies not only a particular
value for $n(k)$ but also that all its
derivatives are equal to zero:
\begin{equation}\label{nn=0bis}
  \frac{d^i n(k)}{d (\ln k)^i}=0, \textrm{ with } i=1,2,...
\end{equation}

Furthermore, the expression for the
${(l_0-1)}^{th}$ derivative of the spectral index contains
slow-roll parameters up to the $l_0^{th}$ one. So once the
approximation order $l_0$ is chosen, the problem is characterized
by $l_0+1$ parameters and $l_0$ equations of constraint relating
them. This allows the expression of \textit{all} the slow-roll
parameters $\lambda_H^l$ as functions of $\epsilon$. The choice of
$\epsilon$ is not arbitrary, because once the expression for
$(\eta-\epsilon)$ appropriate for the approximation level assumed
is derived, the \textit{exact} expressions for $d\epsilon/d\phi$
and for $d\ln V /d\epsilon$, Eqs.~(\ref{depsdphi2}) and (\ref{dlnVdeps}), can be
exploited to compute $\phi$ and $V$ as
functions of $\epsilon$ thus yielding the map $\phi \rightarrow
V(\phi)$.

\acknowledgments{This work was supported in part by NASA grant NAG5-10842.
A.V.~would like to thank the David and Lucile Packard Foundation and Hotel
Victoria, Torino, for financial support.  E.J.C.~thanks the Kavli Institute for
Theoretical Physics, Santa Barbara for their support during the completion of
part of this work.  A.R.L.~was supported in part by the Leverhulme Trust and by
PPARC.  This work was initiated during a visit by E.W.K.~to Sussex
supported by PPARC.  We thank Cesar Terrero-Escalante for extensive comments on
the original
version of this paper, and also Filippo Vernizzi for useful comments.}


\end{document}